\documentclass[prb,superscriptaddress,twocolumn,floatfix]{revtex4-2}

\usepackage{amsmath}
\usepackage{amssymb}
\usepackage[mathscr]{euscript}
\usepackage[hyperindex,breaklinks]{hyperref}
\usepackage{graphicx}
\usepackage{siunitx}
\usepackage{orcidlink}

\DeclareSIUnit\angstrom{\text{\AA}}

\makeatletter
\def\@hangfrom@section#1#2#3{\@hangfrom{#1#2}#3}
\def\@hangfroms@section#1#2{#1#2}
\renewcommand\frontmatter@abstractwidth{\dimexpr\textwidth\relax} \makeatother
\makeatletter

\def\MEMS{\footnotesize Department of Mechanical Engineering and Materials Science, Duke University, Durham, NC 27708, USA}
\def\CAMD{\footnotesize Center for Autonomous Materials Design, Duke University, Durham, NC 27708, USA}
\def\UTDALLAS{\footnotesize Department of Materials Science and Engineering and Department of Chemistry and Biochemistry, University of Texas at Dallas, Richardson, TX 75080, USA}

\def\ZENTRUM{\footnotesize Institute of Ion Beam Physics and Materials Research, Helmholtz-Zentrum Dresden-Rossendorf, 01328 Dresden, Germany}
\def\TCTUD{\footnotesize Theoretische Chemie, Technische Universit\"at Dresden, 01062 Dresden, Germany}
\def\PSU{\footnotesize Department of Materials Science and Engineering, The Pennsylvania State University, University Park, PA 16802, USA}
\def\NCSU{\footnotesize Department of Materials Science and Engineering, North Carolina State University, Raleigh, NC 27695, USA}
\def\BUFFALO{\footnotesize Department of Chemistry, State University of New York at Buffalo, Buffalo, NY 14260, USA}
\def\MST{\footnotesize Department of Materials Science and Engineering, Missouri University of Science and Technology, Rolla, MO 65409, USA}
\def\APLPSU{\footnotesize Applied Research Laboratory, The Pennsylvania State University, University Park, PA 16802, USA}

\begin{document}

\title{A priori procedure to establish spinodal decomposition in alloys}

\author{Simon~Divilov\,\orcidlink{0000-0002-4185-6150}}\affiliation{\MEMS}\affiliation{\CAMD}
\author{Hagen~Eckert\,\orcidlink{0000-0003-4771-1435}}\affiliation{\MEMS}\affiliation{\CAMD}
\author{Cormac~Toher\,\orcidlink{0000-0001-7073-8690}}\affiliation{\UTDALLAS}\affiliation{\CAMD}
\author{Rico~Friedrich\,\orcidlink{0000-0002-4066-3840}}\affiliation{\ZENTRUM}\affiliation{\TCTUD}\affiliation{\CAMD}
\author{Adam~C.~Zettel\,\orcidlink{0000-0003-1645-9476}}\affiliation{\MEMS}\affiliation{\CAMD}
\author{Donald~W.~Brenner\,\orcidlink{0009-0009-1618-4469}}\affiliation{\NCSU}
\author{William~G.~Fahrenholtz\,\orcidlink{0000-0002-8497-0092}}\affiliation{\MST}
\author{Douglas~E.~Wolfe\,\orcidlink{0000-0002-3997-406X}}\affiliation{\APLPSU}
\author{Eva~Zurek\,\orcidlink{0000-0003-0738-867X}}\affiliation{\BUFFALO}
\author{Jon-Paul~Maria\,\orcidlink{0000-0003-3604-4761}}\affiliation{\PSU}
\author{Nico~Hotz\,\orcidlink{0009-0008-2469-2693}}\affiliation{\MEMS}\affiliation{\CAMD}
\author{Xiomara~Campilongo\,\orcidlink{0000-0001-6123-8117}}\affiliation{\CAMD}
\author{Stefano~Curtarolo\,\orcidlink{0000-0003-0570-8238}}\email[]{stefano@duke.edu}\affiliation{\MEMS}\affiliation{\CAMD}

\date{\today}

\begin{abstract}
\noindent
Spinodal decomposition can improve a number of essential properties in materials, especially hardness. Yet, the theoretical prediction of the onset of this phenomenon (e.g., temperature) and its microstructure (e.g., wavelength) often requires input parameters
coming from costly and time-consuming experimental efforts, hindering rational materials optimization.
Here, we present a procedure where such parameters are not derived from experiments.
First, we calculate the spinodal temperature by modeling
nucleation in the solid solution while approaching the spinode boundary.
Then, we compute the spinodal wavelength self-consistently using a few reasonable approximations. Our results show remarkable agreement with experiments and, for NiRh,
the calculated yield strength due to spinodal microstructures surpasses even those of Ni-based superalloys. We believe that
this procedure will accelerate the exploration of the complex materials experiencing spinodal decomposition, critical for their
macroscopic properties.
\end{abstract}

\maketitle

\section{Introduction}
Spinodal decomposition is a state of phase segregation distinguished by a periodic composition modulation that forms after quenching a solid solution inside the spinodal region, a sub-region of the miscibility gap~\cite{Cahn1961795}.
The phenomenon --- characterized by temperature, amplitude and wavelength --- can improve many essential properties of materials such as, hardness~\cite{Cahn_ActaMetall_SpinodalHardening_1963}, magnetic coercivity~\cite{Ditchek_ARMS_1979}, magnetoresistance~\cite{Chen_PRB_1994}, thermoelectric performance~\cite{Androulakis_JACS_2007}, plasticity~\cite{Kim_MMI_2009}, ductility~\cite{Chang_SCRMAT_2010}, and piezoelectric strain coefficients~\cite{Ke_PRL_2020}.

In particular, hardening by spinodal composition occurs because the modulated composition creates obstacles for propagating dislocations by producing periodic stress fields.
Hence, the external force needed to overcome them increases with the amplitude of modulation, and with the density of the fields (Hall–Petch effect) ~\cite{Cahn_ActaMetall_SpinodalHardening_1963,Ardell_MMTRA_1985,Hall_PPS_1951,Petch_JISI_1953,Guo_PRL_B4C_HallPetch_2018}.
Estimating this type of hardness enhancement requires information about the spinodal temperature and wavelength~\cite{Cahn_ActaMetall_SpinodalHardening_1963,Kato_ACTAMETAL_1980}, which are sensitive to the model used to evaluate them.

Modeling spinodal decomposition in alloys has primarily involved \textit{a posteriori} methods, where input parameters need to be extracted from experiments;
e.g., phase-field simulations~\cite{Marro_PRB_1975,Langer_PRA_1975}.
However, this task can be costly and time-consuming.
On the other hand, most \textit{a priori} methods, which do not require experiments, compute only the spinodal temperature based on the cluster expansion formalism~\cite{sanchez_ducastelle_gratias:psca_1984_ce,Teles_PRB_2002,Ferhat_PRB_2002,curtarolo:art107} and Korringa-Kohn-Rostoker theory~\cite{Korringa1947392,Staunton_PRB_1994,Pinski_PRB_1998,Singh_PRB_2015}.
Consequently, predicting hardness enhancement lacks a model where input parameters come from first principles and can calculate both the spinodal temperature and wavelength.

This work introduces an a priori procedure of spinodal decomposition in alloys to determine first, the spinodal temperature and then, the wavelength for hardness enhancement predictions.
We \textbf{i.}\ highlight the approximations and shortcomings of the classical thermodynamic theory of spinodal decomposition;
\textbf{ii.}\ calculate the spinodal temperature from the stability analysis of a nucleation process in the limit of the spinodal (inspired by Binder~\cite{Binder_RPP_1987});
\textbf{iii.}\ self-consistently calculate the spinodal wavelength from a continuous cooling model of spinodal decomposition (derived from Cahn~\cite{cahn_hilliard_1});
\textbf{iv.}\ validate the results with binary alloys where there are reliable experimental measurements~\cite{Massalski} and; due to the importance of rhodium alloy coatings for improving hardness~\cite{Marot_TSF_2008,JACS_Rh_2010},
\textbf{v.}\ predict the hardness enhancement due to the spinodal in the NiRh system, where there is a lack of experimental data~\cite{Nash_BAPD_1984}.

Our procedure enables computational pre-screening and subsequent experimental synthesis for appropriate candidates. The optimization of the separating phases, and the wavelengths~\cite{Hall_PPS_1951,Petch_JISI_1953,Guo_PRL_B4C_HallPetch_2018} can also accelerate the exploration of ultra-hard materials with high-throughput or machine-learning searches~\cite{nmatHT,aflow_software_2022}.
This allows us to guide experiments to synthesize materials within the spinodal region, and with our recently developed AFLOW-Spinodal module, calculate the spinodal wavelength from the electron microscopy~\cite{DEED}.
We anticipate that our methodology will enable the discovery and characterization of systems having spinodal microstructures.

\section{Discussion}
\noindent{\bf Approximations in the classical theory}.
Following the standard thermodynamic treatment of this subject~\cite{cahn_hilliard_1,Hillert1961525}, the excess Helmholtz free energy of a $A_XB_{1-X}$ binary inhomogeneous solid solution is written:
\begin{equation}
    F(X, T) = \int_V \left [ f(x(\mathbf{r}),T) + \frac{\kappa}{V} \left (\nabla x(\mathbf{r}) \right )^2 \right ] d^3\mathbf{r},
    \label{eq:f_classic}
\end{equation}
where $f$ is the bulk excess Helmholtz free energy density, $x$ is the local \textit{microscopic} concentration of species \textit{B}, $\kappa$ is the energy gradient coefficient, $T$ is the temperature and $V$ is the volume of the system.
Here the local concentration is commensurate with the \textit{macroscopic} concentration $X$ so that $X = V^{-1}\int_Vx(\mathbf{r})d^3\mathbf{r}$.
Next, an assumption is made that the volume can be partitioned into sub-volumes $V_i$, and within each sub-volume, $x(\mathbf{r})$ is constrained so that the concentration is $\{ x(\mathbf{r}) = x, \forall \mathbf{r} \in V_i \}$~\cite{Langer_AnnPhys_1971}.
Under this assumption, within each sub-volume, $f(x,T)$ is modeled by a double-well function.
Then, the stability of spinodal decomposition is determined by the sign of $\partial^2 f(x,T) / \partial x^2$, where the temperature at which the second derivative vanishes defines the spinodal temperature~\cite{Hillert1961525}.
Furthermore, the spinodal wavelength is derived as a function of $\partial^2 f(x,T) / \partial x^2$, diverging to infinity at the spinodal temperature~\cite{cahn_hilliard_1}.

Evidently, modeling spinodal decomposition, using the classical theory, requires finding an appropriate functional form for $f$.
For example, in the Bragg-Williams approximation, the energy density is a quadratic function in $x$, while the entropy density is the ideal entropy of mixing, which yields a symmetric double-well potential in concentration space~\cite{Shockley_JCP_1938,cahn_hilliard_1}.
However, as demonstrated by Kikuchi~\cite{Kikuchi_JCP_1967}, improving this approximation results in a lower free energy barrier height.
In fact, using the exact expression for $f$, no free energy barrier exits and no spinodal decomposition occurs because $\partial^2 f / \partial x^2=0$ for the whole concentration space.
Rather, the physical origin of the free energy barrier is due to the quenching procedure, which is present as an additional term in $f$~\cite{Kikuchi_JCP_1967,Binder_PSCA_1986}.
Therefore, using the classical theory of spinodal decomposition would require information about the kinetics or fitting the functional form of $f$ from existing thermodynamic data, which raises insurmountable difficulties for novel or hard-to-synthesize materials.
In the preceding section, we develop a method that does not require a functional form of $f$ or explicit knowledge about the quenching procedure, while also being suitable for our research and first-principles calculations to describe real materials.

\begin{figure}
    \includegraphics[width=0.48\textwidth]{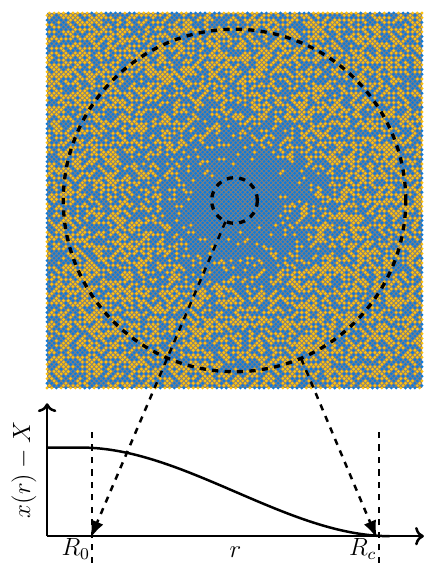}
    \caption{\textbf{Spinodal nucleation model.} Concentration distribution of the spinodal nucleus for a binary equi-composition alloy.
    The variables $R_0$ and $R_c$ is the minimum radius and cut-off, respectively, and are explained in the text.
    \label{fig:model}
    }
\end{figure}

\begin{figure*}
    \includegraphics[width=\linewidth]{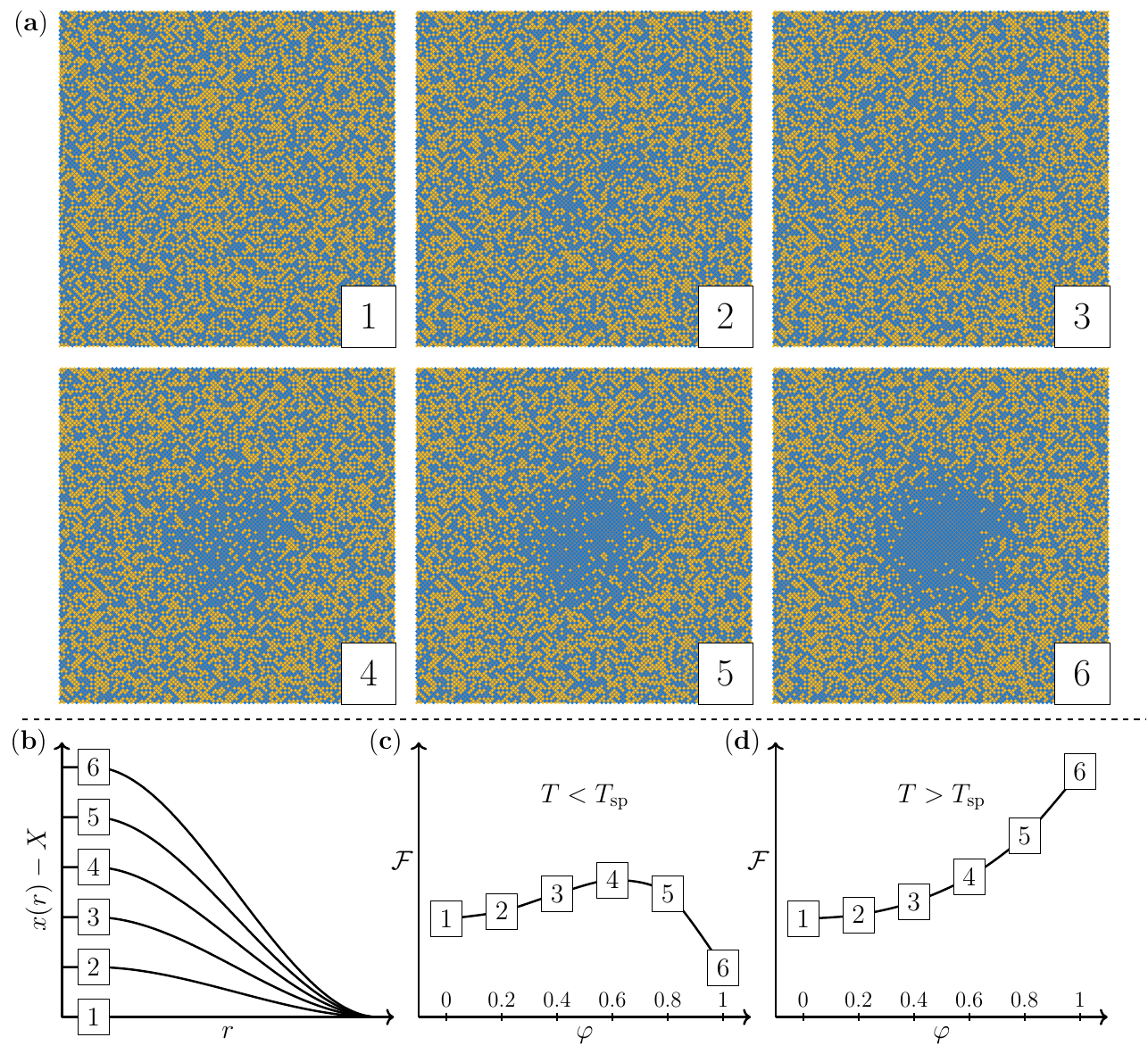}
    \caption{\textbf{Spinodal nucleation stability.} Varying the demixing parameter $\varphi$, for fixed $X$, from the solid solution ($\varphi = 0$) to the maximum phase segregation ($\varphi = 1$) we can determine if spinodal nucleation is stable if the Helmholtz free energy $\mathcal{F}$ of the spinodal nucleus is lower, relative to the solid solution, for increasing $\varphi$.
    For a given $\varphi$, (\textbf{a}) the image of atoms shown is representative of the (\textbf{b}) atomic distributions.
    $\mathcal{F}(X, T, \varphi)$ is calculated by averaging the Helmholtz free energy of many such configurations.
    (\textbf{c}-\textbf{d}) When the temperature $T$ is below (above) $T_\text{sp}$, $\mathcal{F}(\varphi=1)$ is lower (higher) than the solid solution $\mathcal{F}(\varphi=0)$ and spinodal nucleation is stable (unstable).
    \label{fig:demix}
    }
\end{figure*}

\noindent{\bf A priori procedure.} Our strategy to model spinodal decomposition, without any external parameters, is to consider a nucleation event in the solid solution infinitesimally close to the spinodal boundary --- i.e., spinodal nucleation.
Since it has been shown that successful spinodal nucleation should precede spinodal decomposition~\cite{Binder_PRA_1984,Unger_PRB_1984,Binder_RPP_1987}, we avoid simulating the concentration modulation explicitly, and with a few reasonable assumptions, still evaluate the spinodal wavelength.

The nucleus in the alloy is characterized by a diffuse radial interface between the emerging segregated phase, in the center of the nucleus, and the bulk phase, described by a solid solution~\cite{cahn_hilliard_2}.
It is shown in Fig.~\ref{fig:model} with the concentration profile generally represented by a logistic function~\cite{Monette_JStatPhys_1992,Monette_PRL_1992} and approximated using a cubic spline:
\begin{widetext}
    \begin{equation*}
        x(\mathbf{r}=r)-X =
        \begin{cases}
            \Delta x & r \leq R_0\\
            \Delta x \left [ 1 - 3\left ( \frac{r - R_0}{R_c - R_0} \right )^2 + 2\left ( \frac{r - R_0}{R_c - R_0} \right )^3 \right ] & R_0 < r < R_c\\
            0 & r \geq R_c
        \end{cases},
    \end{equation*}
\end{widetext}
where $\Delta x \in [-\min(X,1-X), \min(X,1-X)]$ is the concentration amplitude, $R_0$ is the minimum radius, and $R_c$ is the cut-off to account for the periodic boundary conditions of the simulation.
The minimum radius is required to properly perform the discretization of the concentration profile over a continuous variable for an atomistic model.

The successful evolution of spinodal nucleation consists of \textit{growing} nuclei, identified by arbitrarily large extents ($R_c \gg R_0$) and concentration amplitudes~\cite{Unger_PRB_1984}.
Therefore, the stability of nuclei with such features --- defined by an excess Helmholtz free energy $\mathcal{F}$ lower than the solid solution --- can serve as an indicator for the stability of nucleation.
To perform our stability assessment, outlined in Fig.~\ref{fig:demix}, we define a demixing parameter $\varphi = \pm\Delta x/\min(X,1-X)$, so that $\varphi=0$ for the solid solution and $\varphi=\pm1$ at the maximum concentration amplitude, the final state of the growing nucleus.
Then, varying $\varphi$ and calculating the corresponding $\mathcal{F}$, we determine if nucleation is stable (unstable) if the growing nucleus is progressively lower (higher) in energy than the solid solution.
Since the alloy is defined by the random occupation of the lattice sites, we work with the space of possible configurations $\Omega(X, \varphi)$ generated by the demixing process for a given macroscopic concentration.
Hence, the proper technique~\cite{Langer_ACTAMETAL_1973} to calculate $\mathcal{F}$ is to perform an ensemble average by:
\begin{equation}
    \mathcal{F}(X, T, \varphi) = -k_BT\log \left \{ \sum_{\{x\}\in \Omega(X, \varphi)}\!\!\!\!\!\exp\left [ \frac{-F(\{x\}, T)}{k_BT} \right ] \right \}
    \label{eq:f_average},
\end{equation}
where $F$ is the excess Helmholtz free energy of a particular configuration and $k_B$ is the Boltzmann constant.
We emphasize that spinodal nucleation is stable \textit{only if} $\mathcal{F}(\phi = 1) < \mathcal{F}(\phi = 0)$, and not based on the sign of $\partial^2\mathcal{F}/\partial\varphi^2$, as would be expected from a naive comparison to spinodal decomposition.

It has been shown~\cite{Binder_PRA_1984,Unger_PRB_1984,Binder_RPP_1987} that the stabilities of both phenomena are logically equivalent; if the former is stable, so is the latter.
In addition, the spinodal temperature of the stability $T_\text{sp}$, in both cases, is approximately equal (see Ref.~\cite{Binder_RPP_1987} for a detailed discussion).
Thus, we can postulate that the temperature $T_\text{sp}(X)$ of both spinodal nucleation and spinodal decomposition is given by the locus:
\begin{equation*}
    \mathcal{F}(X, T_\text{sp}, \varphi=1) - \mathcal{F}(X, T_\text{sp}, \varphi=0) = 0,
\end{equation*}
which is similar to the Kullback-Liebler entropy-loss locus
in the Lederer-Toher-Vecchio-Curtarolo method to calculate the miscibility gap~\cite{curtarolo:art139}.
To find such locus, we evaluate Eq.~(\ref{eq:f_average}), and compute the excess Helmholtz free energy as:
\begin{equation*}
    F(\{x\}, T) = \left [U(\{x\}) - U_\text{rnd}(X) \right ] - T \left [S(\{x\}) - S_\text{rnd}(X) \right ],
\end{equation*}
where $U$ is the internal energy of the state, $S$ is the configurational entropy of the state, and
\begin{gather*}
        U_\text{rnd}(X) = \frac{1}{|\Omega(X, 0)|}\sum_{\{x\}\in \Omega(X, 0)}U(\{x\}),\\
        S_\text{rnd}(X) = -k_B\left [X\log X + (1-X)\log(1-X) \right ]
\end{gather*}
are the internal energy and configurational entropy of the solid solution, respectively.
The evaluation of $U$ depends on the model interaction, while $S$, for arbitrary configurations, is computed following Ref.~\cite{Araki_COMMMP_1970}.

However, the calculation of the spinodal wavelength is more complicated and is given by the self-consistent solution to the equation~\cite{Evans_AP_1979}:
\begin{equation*}
    \hat{c}(X, T, \lambda_\text{sp}) = 1,
\end{equation*}
where $\hat{c}$ is the Ornstein-Zernike direct correlation function~\cite{Stell_PRB_1970} of the concentration modulation relative to the solid solution and $\lambda_\text{sp}$ is the spinodal wavelength.
Assuming that the length scale of the concentration modulation is large relative to the lattice constant of the material, as is the case in our atomic configurations, $\hat{c}$ can be expanded in negative powers of $\lambda_\text{sp}$.
Then, equating the coefficients with the gradient expansion of the excess free energy up to second order~\cite{cahn_hilliard_1,Evans_AP_1979}, we arrive at:
\begin{equation*}
    4\pi^2\kappa(X)\lambda^{-2}_\text{sp}(X,T) + \frac{1}{2}\frac{\partial^2 f\left (x(\mathbf{r}),T\right )}{\partial x^2}\Bigg|_{x=X} = 0,
\end{equation*}
where $f$ and $\kappa$ are the same quantities as in Eq.~(\ref{eq:f_classic}).
Next, we avoid defining a functional form for $f$ by considering spinodal nucleation to occur under continuous cooling so that the Taylor series of the excess free energy density is truncated to first order in temperature~\cite{Huston_ACTAMETAL_1966}.
The first-order coefficient is just the excess entropy density, and, for crystalline systems, it can always be well approximated by the ideal entropy~\cite{Cadoret_PSSB_1971}.
This yields a simplified formula for the spinodal wavelength:
\begin{equation}
    \lambda_\text{sp}(X, T) = \sqrt{\frac{-16\pi\kappa(X)X(1-X)}{k_B\left (T-T_\text{sp}(X)\right )}}.
\end{equation}
\begin{figure*}
    \includegraphics[width=\linewidth]{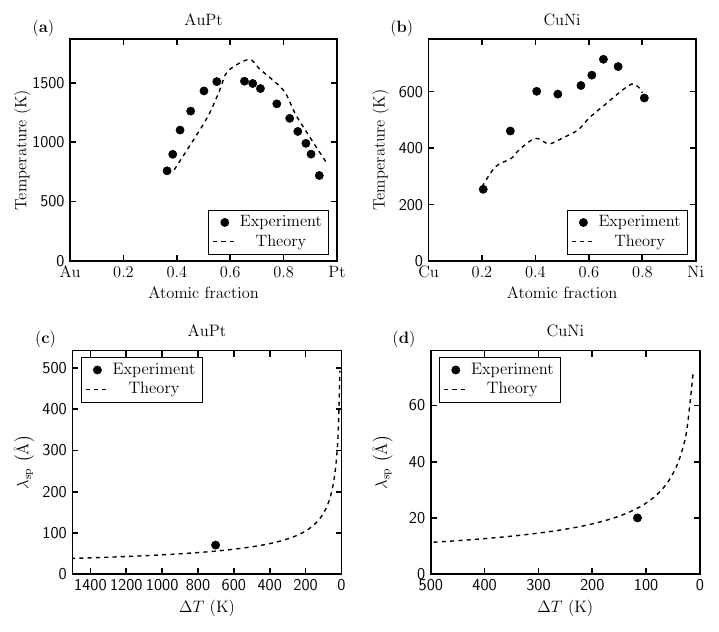}
    \caption{\textbf{Comparison between theory and experiment.}
             (\textbf{a}-\textbf{b}) Spinodal temperature, as a function of composition, for AuPt and CuNi.
             (\textbf{c}-\textbf{d}) Maximum spinodal wavelength, as a function of the undercooling temperature, for 40/\SI{60}{at{.}\%} Au/Pt and 30/\SI{70}{at{.}\%} Cu/Ni alloys.
    }
    \label{fig:sp_comparison}
\end{figure*}

Furthermore, if the crystallographic point group of the alloy has inversion symmetry and the model interaction does not depend on the local concentration, we have (see Appendix for further details):
\begin{equation}
    \kappa(X) = \frac{1}{6}\sum_nZ_n\nu(r_n(X))r_n(X)^2,
    \label{eq:kappa}
\end{equation}
where
\begin{equation*}
    \nu(r) = \phi^{12}(r) - \frac{1}{2}\left [\phi^{11}(r) + \phi^{22}(r) \right ].
\end{equation*}
Here $Z_n$ and $r_n$ are the coordination number and radius  of the $n$th coordination shell, and $\phi^{ij}$ is the interatomic potential between component $i$ and $j$.
For crystal structures with lower symmetry, it is possible to decompose the system into sub-lattices~\cite{Hillert_ACTAMETAL_1977}, such that $\kappa(X) = \sum_\alpha\kappa_\alpha(X)$, where $\alpha$ runs over the sub-lattices indices.
Undoubtedly, $U$ and $\nu$ must be evaluated with the same model interaction so that $T_\text{sp}$ and $\lambda_\text{sp}$ are consistent.

The atomistic model we have developed naturally allows us to simulate the interactions in the alloy using first-principles or interatomic potentials techniques.
To calculate the spinodal using the present theory, only the interatomic potential needs to be determined, which, following the literature, can be generated directly using density functional theory calculations~\cite{curtarolo:art174}.
Since spinodal nucleation is a morphological phenomenon~\cite{Mecke_PRE_1997} characterized by the spatial variation of the concentration in the interface, we expect the method to be quite insensitive to the details of the interatomic potential model.

\section{Results}

\begin{figure*}
    \includegraphics[width=\linewidth]{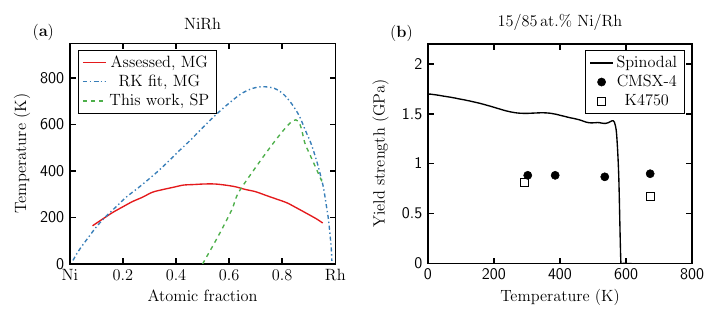}
    \caption{\textbf{Properties of NiRh.}
             (\textbf{a}) Phase diagram as a function of temperature and concentration. The miscibility gap (MG) was calculated by an assessment of the experimental data~\cite{Nash_BAPD_1984} and also a Redlich-Kister (RK) fit of ab initio calculations~\cite{Teeriniemi_JAC_2015}.
             The spinodal (SP) was calculated using the technique developed in this work.
             (\textbf{b}) Contribution to the yield strength from spinodal decomposition, in the 15/\SI{85}{at{.}\%} Ni/Rh alloy, as a function of temperature, assuming mixed dislocations.
             CMSX-4 and K4750 are Ni-based superalloys~\cite{Sengupta_CMSX4_1994,Hou_JAC_2022}.
    }
    \label{fig:ni_rh}
\end{figure*}

We have applied our a priori methodology to two binary alloys, AuPt and CuNi.
The two systems were chosen because there existed ample experimental data on the spinodal temperature~\cite{Okamoto_BAPD_1985,Vrijen_PRB_1978} and measurements of the spinodal wavelength for select concentrations~\cite{Singhal1978_JACRYST_1978,AvilaDavila_JAC_2008}.
The excess free energy was calculated with LAMMPS~\cite{LAMMPS_2022}, with the energy minimization performed using the steepest decent algorithm, due to its robustness.
The interatomic potentials, for both systems, taken from the literature~\cite{OBrien_JMS_2017,Fischer_ACTAMAT_2019}, were generated based on the embedded atom method~\cite{Daw_PRB_1984}.
For each concentration, we determined the equilibrium lattice constant of the random solution with a fit to the Murnaghan equation of state~\cite{Murnaghan_PNAS_1944}, which was kept fixed throughout the simulations for both the initial and terminal states.
Then, we computed $\mathcal{F}$ of an ensemble of $|\Omega| = 100$ configurations and the spinodal temperature was found using Brent's method.
The converged geometric parameters were determined to be $R_0 \approx$ \SI{1}{\nano\meter} and $R_c \approx$ \SI{10}{\nano\meter}.
These parameters are independent of the species and composition, which is further evidence that we are really simulating a morphological process~\cite{Mecke_PRE_1997}.

The comparison of $T_\text{sp}$, between the theory and experiment, shown in Fig.~\ref{fig:sp_comparison}, presents remarkable agreement.
The slight under-approximation of $T_\text{sp}$ for CuNi, could be due to the nature of the experiment, which required the alloy to be enriched with heavier isotopes of Cu and Ni~\cite{Vrijen_PRB_1978}.
We have also recomputed the CuNi spinodal using another interatomic potential from the literature~\cite{Onat_JPCM_2013} and found similar results.

Furthermore, the calculated maximum spinodal wavelength, shown in Fig.~\ref{fig:sp_comparison}, has the expected dependence on the undercooling temperature $\Delta T \equiv T_\text{sp}-T$ --- diverging at the spinodal temperature~\cite{cahn_hilliard_1}.
The comparison with the experimental data, while limited, is in agreement with our calculations.
The $\kappa$ was found to be \SI{4.953}{eV\angstrom^2} and \SI{0.166}{eV\angstrom^2} for 40/\SI{60}{at{.}\%} Au/Pt and 30/\SI{70}{at{.}\%} Cu/Ni, respectively.
Relative to the $\kappa$ obtained from the regular solution model~\cite{cahn_hilliard_2}, which only depends on the nearest-neighbor distance and $T_\text{sp}$, these values were approximately, a factor of three, larger and smaller, for AuPt and CuNi, respectively.
This inconsistency strongly suggests that $\lambda_\text{sp}$ cannot be accurately captured with a theory that only predicts $T_\text{sp}$, but also requires input about the local energy landscape of the alloy.

Next, we extended our approach to the NiRh system, which has numerous applications as a hardness coating and alloying agent~\cite{JACS_Rh_2010}.
In addition, recent theoretical work~\cite{Teeriniemi_JAC_2015} reveals a different phase diagram than the one obtained from experimental assessment~\cite{Nash_BAPD_1984}.
Our results of the spinodal, obtained using an angular dependent interatomic potential~\cite{Mishin_ACTAMAT_2005,Xu_JSCC_2022}, are consistent with the miscibility gap obtained by the Redlich-Kister (RK)~\cite{Redlich_IndsEngChem_1948} fit of ab initio calculations (see Fig.~\ref{fig:temp_nirh} for the comparison).
Similar to the calculated miscibility gap, we observed a maximum in the spinodal at high concentrations of Rh, while no spinodal decomposition occurs beyond \SI{50}{at{.}\%} Ni, near the onset of a ferromagnetic phase transition~\cite{Muellner_PRB_1975}.

Moreover, we have computed the contribution to the yield strength from spinodal decomposition at the critical point (\SI{15}{at{.}\%} Ni), shown in Fig.~\ref{fig:ys_nirh}, by numerically solving the boundary value problem of the force balance equation, for the mixed dislocations, following Ref.~\cite{Kato_ACTAMETAL_1980}.
The elastic constants used in the equation were evaluated using Vegard's law from experimental data of the constituent metals~\cite{Salama_PSSA_1977,Maurer_PSSA_1997}.
The yield strength surpasses that of the Ni-based superalloys~\cite{Sengupta_CMSX4_1994,Hou_JAC_2022}, and is relatively constant up to \SI{550}{K}, but begins to rapidly fall off at higher temperatures.
This is because the spinodal wavelength is diverging, and thus, a larger quantity of dislocations can pile up near the boundary of the concentration modulation, requiring a smaller driving force for the dislocations to propagate through the material~\cite{Cahn_ActaMetall_SpinodalHardening_1963}.
Nonetheless, this result illustrates the possibility of engineering materials whose yield strength exceeds those of high-performance alloys, particularly when the spinodal temperature is high.

\section{Conclusions}
Spinodal decomposition is a promising, but often neglected, phenomenon that can be exploited to increase an alloy's hardness.
However, its use has been mostly limited because a comprehensive characterization of spinodal decomposition requires experimental input.
Here we have developed an a priori procedure, based on the stability analysis of an intermediary process.
Effectively, we represent the spinodal decomposition as a late-stage spinodal nucleation, which is straightforward to simulate.
Then, using a model derived from Cahn, we self-consistently compute the spinodal wavelength.
Only the interatomic potential between the species is necessary to perform these calculations.

We have applied our procedure to well-explored alloys and have reproduced experimental results with impressive accuracy.
In addition, we have made predictions for the NiRh alloy, where experiments were absent.
Our results are consistent with the miscibility gap obtained from first-principles calculations, and we have found that the yield strength due to spinodal decomposition exceeds those of high-performance alloys.
We are currently working to extend this idea to multi-component alloys, and predict that this work will
be valuable to understand hardness enhancement by spinodal decomposition.

\section*{Declaration of Competing Interest}
The authors declare that they have no competing financial interests
or personal relationships that could have appeared to influence the work
reported in this paper.

\section*{Acknowledgments}
  The authors thank Ohad Levy, Arrigo Calzolari, Aaron Stebner and Sean Griesemer
  for fruitful discussions.
  Funding for this research was provided by the Office of Naval
  Research through a Multidisciplinary University Research Initiative
  (MURI) program under project number N00014-21-1-2515,
  N00014-23-1-2615, and by the National Science Foundation (NSF NRT-HDR DGE-2022040).
  This work was supported by high-performance computer time and resources from the DoD
  High Performance Computing Modernization Program (Frontier).
\\

 \section*{Appendix}
For a lattice with inversion symmetry, occupied by two species, $A$ and $B$, it was shown that~\cite{Hildebrand_JCP_1933,cahn_hilliard_1}:
\begin{equation*}
    \kappa^{(1)}(X) = -\frac{X}{6}\sum_nZ_nr_n^2(X)\nu(X, r_n(X)),
\end{equation*}
where $Z_n$ and $r_n$ is the coordination number and radius of the $n$th coordination shell, and $\nu$ is the pair-wise interaction.
The energy gradient coefficient is given by $\kappa = -\partial\kappa^{(1)}/\partial X$~\cite{cahn_hilliard_1}, which yields:
\begin{equation*}
    \begin{split}
        \kappa &= \frac{1}{6}\sum_nZ_nr_n^2\nu +
                \frac{X}{3}\sum_nZ_nr_n\frac{\partial r_n}{\partial X}\nu + \\
                & + \frac{X}{6}\sum_nZ_nr_n^2\left [\frac{\partial \nu}{\partial r_n}\frac{\partial r_n}{\partial X} + \frac{\partial \nu}{\partial X} \right ],
    \end{split}
\end{equation*}
where we have dropped the dependence on $X$ for brevity.
Next, if we assume that $\nu$ is concentration independent and that Vegard's law holds true --- $r_n(X) \equiv (1-X)r^{A}_n + Xr^{B}_n$ --- we have:
\begin{equation*}
    \begin{split}
        \kappa &= \frac{1}{6}\sum_nZ_nr_n^2\nu
                 + \frac{X}{3}\sum_nZ_nr_n\nu\delta_n + \\
                & + \frac{X}{6}\sum_nZ_nr_n^2\frac{\partial \nu}{\partial r_n}\delta_n,
    \end{split}
\end{equation*}
where $\delta_n \equiv r^{B}_n - r^{A}_n$.
Evidently, $\{ n\in \mathbb{N} \mid \delta_n \ll r_n \}$, and we can reasonably drop the terms with $\delta_n$ to arrive at Eq.~(\ref{eq:kappa}).

\newcommand{\Ozolins}{Ozoli{\c{n}}{\v{s}}}

\end{document}